\documentclass[aps,prb,twocolumn,showpacs,letterpaper,superscriptaddress]{revtex4-1}

\usepackage[colorlinks=true, citecolor=blue, linkcolor=blue, urlcolor=blue]{hyperref}
\usepackage{graphicx,dcolumn,longtable,epsfig}
\usepackage[usenames]{color}
\usepackage{amssymb}
\usepackage{amsmath}
\usepackage{bm}
\usepackage{footnote}
\usepackage{float}
\usepackage{subfigure}
\usepackage{color}

\usepackage{ulem}
\usepackage[T1]{fontenc}

\newcommand{\be}{\begin{equation}}
\newcommand{\ee}{\end{equation}}

\def\bea{\begin{eqnarray}}
\def\eea{\end{eqnarray}}

\begin{document}

\title{Light polarons with electron-phonon coupling}

\author{Chao Zhang}
\email{zhangchao1986sdu@gmail.com}
\affiliation{Department of Modern Physics, University of Science and Technology of China, Hefei, Anhui 230026, China}
\affiliation{Hefei National Laboratory, University of Science and Technology of China, Hefei, Anhui 230088, China}

\begin{abstract}

In most cases, as the strength of electron-phonon coupling increases, the effective mass of polarons typically increases. However, in this paper, we uncover a fascinating phenomenon: the presence of light polarons even within the strong coupling regime, where electron-phonon coupling includes both Holstein and bond types (electron-phonon coupling occurring on the hopping bonds). To investigate this, we employ a novel Diagrammatic Monte Carlo method based on the path-integral formulation of the particle sector and the Fock-state representation of the phonon sector. Our study centers on analyzing the impact of bond electron-phonon coupling on the Holstein polaron's essential properties, including its effective mass, ground state energy, and the average number of phonons. We examine two distinct scenarios: one where the phonon frequency of both Holstein and bond types is the same, and another where the phonon frequency of the Holstein type is twice as large as that of the bond type. In both of these two cases, when the Holstein coupling falls within the light mass regime, we observe minimal changes in the effective mass as a function of bond coupling, compared to the case of the bare bond polaron. However, the behavior of the effective mass undergoes a significant change in the heavy mass regime. Most intriguingly, we discover a non-monotonic dependence of the effective mass on the bond coupling $g_B$ when the Holstein coupling $g_H$ falls within the heavy mass regime. This finding holds promise, particularly in the context of bipolaron, where achieving a light effective mass and a large binding energy to get the compact size, is crucial.

\end{abstract}

\maketitle

\section{Introduction}

Electron-phonon coupling is a fundamental interaction in condensed matter physics~\cite{Landau33,Frohlich50,Feynman:1955du, Schultz:1959el, Holstein59, Alexandrov:1999fy, Holstein2000}. It represents the coupling between the motion of electrons (charge carriers) and the vibrational modes of the crystal lattice (phonons) within a solid-state material. The electron-phonon coupling leads to the formation of polarons~\cite{}, where electrons induce localized lattice distortions, and plays a pivotal role in the emergence of superconductivity when bipolarons are formed~\cite{PhysRevX.13.011010, Johnarxiv2022}. 
The bipoaron formed by electron-phonon coupling is intricately linked to the mechanism of high-temperature superconductivity in the dilute-density limit. In dilute-density limit, the electron-phonon interaction has the capacity to bind two polarons into a singular bipolaron, akin to the formation of a Bose-Einstein condensate-like superconductor. For such a superconducting state to exists, specific prerequisites, including a bipolaron with a light effective mass, compact size, and a large phonon-mediated pairing potential, must be met.~\cite{PhysRevX.13.011010,Johnarxiv2022, PhysRevLett.130.236001}. 

Depending on whether the phonon vibrations are coupled to the electron density or the hopping motion of electrons, two primary types of electron-phonon coupling come into play: the Holstein type and bond Su-Schrieffer-Heeger (bond) type. Extensive prior research has revealed that in the Holstein model, where the electron-phonon coupling predominantly influences the electron density, the effective mass of both polaron and bipolaron exhibits exponential growth at strong electron-phonon coupling strengths~\cite{PhysRevB.55.R8634, PhysRevLett.105.266605, PhysRevLett.84.3153, PhysRevB.69.245111, PhysRevLett.81.433, PhysRevB.43.2712, Costa_1996, PhysRevB.54.12835}. In contrast, the scenario alters significantly when considering the bond polaron, where the electron-phonon coupling couples to the hopping of electrons. Recent investigations into bond polarons have garnered substantial interest, chiefly due to the remarkable characteristic of a non-exponentially effective mass even in strong coupling regime, resulting in a light polaron and bipolaron~\cite{PhysRevLett.105.266605, PhysRevB.104.035143, PhysRevB.104.L140307}, a feature with significant implications for understanding electron-phonon couplings in advanced materials and quantum systems.

In this paper, we investigate the effects of the bond electron-phonon coupling on the properties of the Holstein polaron in a two dimensional square lattice using a newly developed Diagrammatic Monte Carlo (DiagMC) method. Our approach is based on the path-integral formulation of the particle sector combined with Fock path-integral representation for the phonon sector~\cite{PhysRevB.105.L020501}. To the best of our knowledge, our calculation is the first quantitative work to (i) demonstrate, using an unbiased approach, the properties of polaron with both of these two electron-phonon couplings and (ii) study the properties of two distinguished phonon frequencies $\omega_H/t=\omega_B/t$ and $\omega_H/t=2\omega_B/t$ in the adiabatic regime $\omega_H/t \le 1.0$ and $\omega_B/t \le1.0$. 
Most importantly, we find that in both of these situations, when the Holstein coupling is situated in the light mass regime, we observe minimal changes in the effective mass as a function of the bond coupling, denoted as $g_B$, compared to the scenario of the bare bond polaron. However, a significant change in the behavior of the effective mass emerges as we enter the heavy mass regime of the Holstein polaron. What's particularly intriguing is the discovery of a non-monotonic relationship between the effective mass and the bond coupling strength $g_B$ when the Holstein coupling $g_H$ falls within the heavy mass regime. The rest of the paper is organized as follows. In Sec.~\ref{sec:sec2}, we present the Hamiltonian of the polaron with both Holstein and bond electron-phonon couplings. In Sec.~\ref{sec:sec4}, we revisit the main properties of the Holstein polaron in the adiabatic regime. In Sec.~\ref{sec:sec5}, we discuss the results, and Sec.~\ref{sec:sec6} concludes the paper.

\section{Hamiltonian}
\label{sec:sec2}

We investigate a polaron model on a two-dimensional square lattice, which has both Holstein and bond electron-phonon couplings. In this model, the electron-phonon coupling consists of two distinct components: (i) the coupling between the electron's density and phonons, which is the Holstein type~\cite{PhysRevB.65.174306, PhysRevB.60.1633}, and (ii) the coupling between the electron's hopping and phonons, which is the bond type. The second component of the bond type electron-phonon coupling implies that the electronic hopping between two lattice sites is influenced by a single oscillator located on the bond connecting these two sites~\cite{PhysRevLett.42.1698, PhysRevLett.25.919, PhysRevB.5.932, PhysRevB.5.941}. The Hamiltonian that characterizes this model is expressed as:

\begin{equation}
\begin{aligned}
& H=H_e+H_{ph}+H_{int} \\
& H_e=-t\sum_{\langle i j \rangle, \sigma} (c_{j,\sigma}^{\dagger} c_{i,\sigma} + H.c.)  \\
& H_{ph}=\omega_B \sum_i (b_i^{\dagger}b_i + 1/2) + \omega_H \sum_{i} (b_i^{\dagger}b_i +1/2) \\
& H_{int}=g_H \sum_i c_i^{\dagger} c_i X_i +g_B \sum_{\langle i j \rangle, \sigma} (c_{j,\sigma}^{\dagger} c_{i,\sigma} +H.c.) X_{\langle ij \rangle} 
\label{Eq1}
\end{aligned}
\end{equation}

Here, the Hamiltonian of the system contains three parts: the electron part, the phonon part and the electron-phonon interaction part. In the electron part, the electron has kinetic energy and it can hop to its nearest neighboring sites. Here, $\langle i j \rangle$ denotes the nearest-neighbor sites, and $t$ is the electron hopping amplitude between the nearest-neighbor sites (we use it as the unit of energy). $c_{i,\sigma}$ ($c^{\dagger}_{i,\sigma}$) is the electron annihilation (creation) operators on site $i$ with spin $\sigma \in \{\uparrow, \downarrow \}$. The second part is the phonon part and $b_i$ ($b^{\dagger}_i$) is the phonon annihilation (creation) operators on site $i$ and $\omega_H$ is the phonon frequency of the Holstein coupling and $\omega_B$ is the phonon frequency of the bond coupling. The third part represents the electron-phonon interaction. $g_H$ is the electron-phonon coupling strength for the Holstein type with  $X_i=b_i+b_i^{\dagger}$ as the oscillator associated with site $i$. $g_B$ is the electron-phonon coupling strength of the bond type with $X_{\langle ij \rangle}=b_{\langle ij \rangle} + b_{\langle ij \rangle}^{\dagger}$ as the oscillator associated with the bond connecting site $i$ and $j$.

The methodology we use is the DiagMC method, which is well-established in solving a wide range of polaron and bipolaron problems~\cite{PhysRevLett.81.2514, PhysRevB.62.6317, PhysRevLett.123.076601, PhysRevLett.105.266605, PhysRevLett.113.166402, PhysRevB.106.L041117}. Here, DiagMC is based on the path-integral formulation of the particle sector and the Fock state representation of the phonon sector. Detailed information about this method can be found in REF~\cite{PhysRevB.105.L020501}. 

The properties of the polaron is controlled by the adiabaticity ratio $\omega_H/t$ and $\omega_B/t$. Here, in this paper, we work on the adiabatic regime $\omega_H/t \le 1.0$ and $\omega_B/t \le 1.0$, where the phonon degree of freedom is considered comparable or slow with respect to the electron motion. 

\begin{figure}[t]
\includegraphics[width=0.45\textwidth]{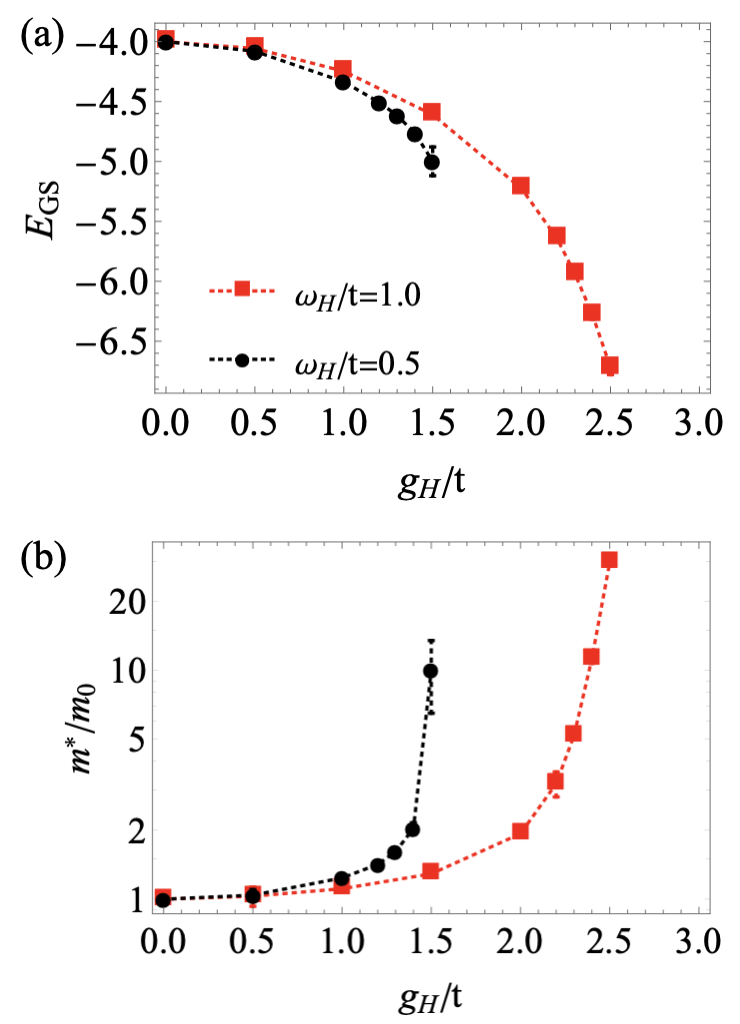}
\caption{The properties of the Holstein polaron: the ground state energy $E_{GS}$ (a), the effective mass $m^*/m_0$ with $m_0=1/2ta^2$ as the mass of one electron (b), as a function of electron-phonon coupling $g_H/t$ with the Holstein phonon frequency $\omega_H/t=0.5$ (black dots) and 1.0 (red squares). If not visible, error bars are within symbol size. 
} 
\label{FIG1}
\end{figure}

\section{Holstein polaron}
\label{sec:sec4}

In the Holstein model, the electron-phonon coupling represents the interaction between phonons and the electron density. This coupling arises from localized lattice distortions induced by electrons to accommodate their presence. It is widely agreed that, as the strength of electron-phonon coupling increases, there is an exponential growth in the effective mass of Holstein polaron~\cite{PhysRevB.65.174306}. This effect is particularly prominent within the regime of strong electron-phonon coupling, where Holstein polaron exhibit a substantial increase in it effective mass, leading to significant implications for charge transport properties. Moreover, when two polarons form a bound state, resulting in a bipolaron, it is noteworthy that the bipolaron also exhibits an exponential increase in its effective mass within the strong electron-phonon coupling regime~\cite{PhysRevB.69.245111}. The heavy effective mass further underscores the possibility of the Holstein model for describing high-temperature superconductivity, as such heavy effective masses are incompatible with the conditions required for high-temperature superconductivity to occur.

In this section, we revisit the properties of Holstein polaron, specifically its ground state energy and effective mass, as a function of the electron-phonon coupling parameter denoted as $g_H/t$, while keeping the Holstein phonon frequency fixed at $\omega_H/t=0.5$ and $\omega_H/t=1.0$. Notably, the Holstein polaron has been extensively investigated in the literature, and we present these results here for two main purposes: (i) to provide essential context for the subsequent discussion in Section~\ref{sec:sec5}, where we introduce bond coupling $g_B/t$ to the Holstein polaron, and (ii) to serve as a benchmark for the methodology employed in our study. Importantly, our results obtained through the DiagMC method is consistent with the results established in literature~\cite{PhysRevB.65.174306}.

Figure~\ref{FIG1} presents the ground state energy (a) and the effective mass (b) of the Holstein model with phonon frequencies $\omega_H/t=0.5$ and 1.0, respectively, as a function of the Holstein coupling parameter $g_H/t$. It can be seen that the ground state energy $E_G$ exhibits a gradual decrease as the coupling strength increases, irrespective of the specific phonon frequency $\omega_H/t$. Moreover, for a given electron-phonon coupling $g_H/t$, the ground state energy is significantly lower in scenarios featuring a smaller phonon frequency. Figure~\ref{FIG1}(b) shows the effective mass of the Holstein polaron as a function of the coupling parameter $g_H/t$ for these two phonon frequencies. Remarkably, a crossover occurs from a light polaron state to a heavy polaron state in both cases, with the crossover taking place at approximately $g_H/t \sim 1.2$ for $\omega_H/t=0.5$ and $g_H/t \sim 2.2$ for $\omega_H/t=1.0$. Furthermore, it is evident that the effective mass increases as the phonon frequency decreases at a constant coupling strength. The crossover from a light polaron state to a heavy polaron state in the Holstein model establishes the foundation for the following discussion in Section~\ref{sec:sec5}.

\begin{figure}[t]
\includegraphics[width=0.42\textwidth]{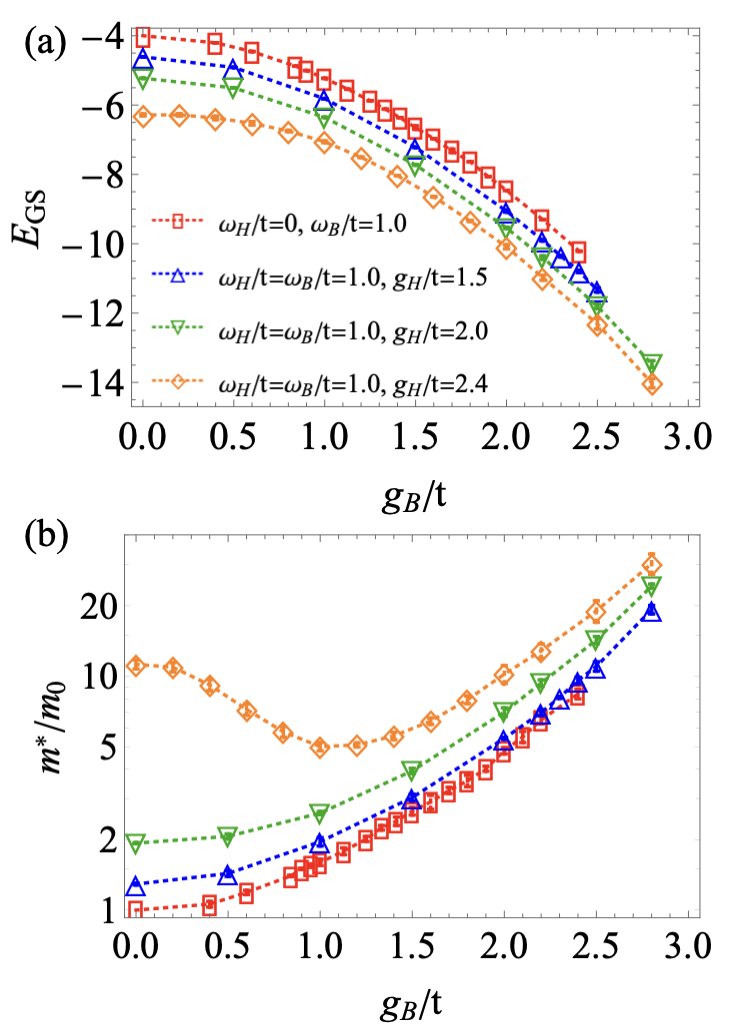}
\caption{The figures presented herein correspond to the phonon frequency $\omega_H/t=\omega_B/t=1.0$. The properties of the Holstein polaron: the ground state energy $E_{GS}$ (a), the effective mass $m^*/m_0$ (b), at electron-phonon coupling strength $g_H/t=1.5$ (blue up triangles), 2.0 (green down-triangles), and $2.4$ (orange diamonds), as a function of the bond electron-phonon coupling $g_B/t$. The properties of the bond polaron as a function of the bond electron-phonon coupling $g_B/t$ at the bond phonon frequency $\omega_B/t=1.0$ are also presented here for comparison (red rectangles). If not visible, error bars are within symbol size. 
} 
\label{FIG2}
\end{figure}

\begin{figure}[t]
\includegraphics[width=0.41\textwidth]{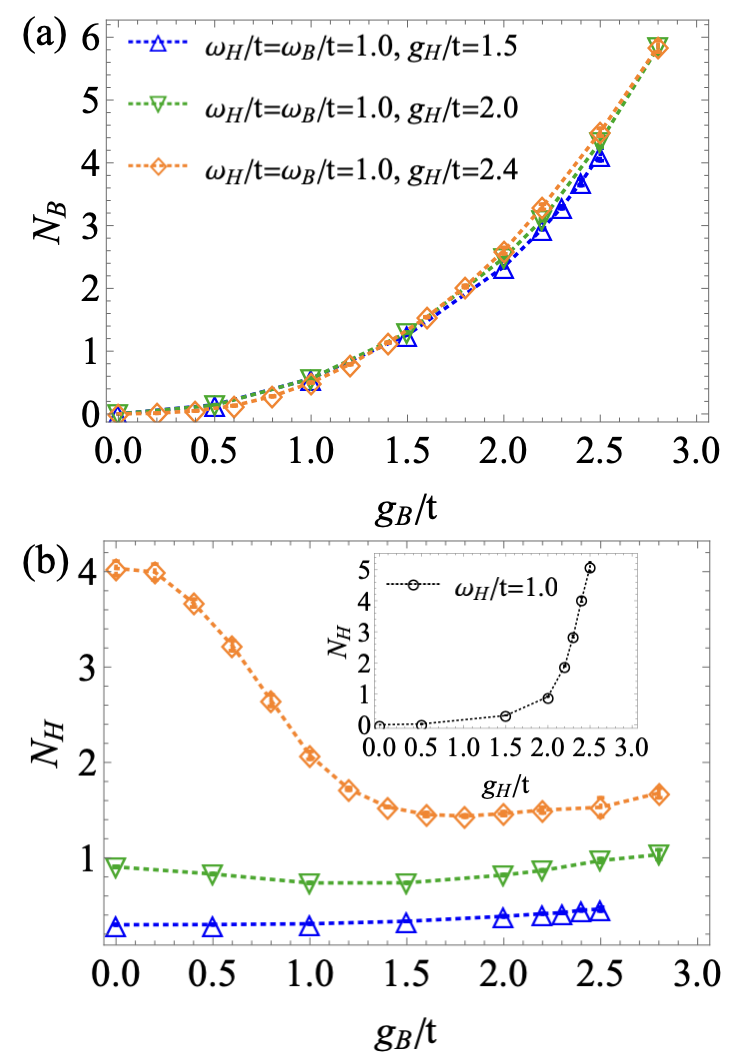}
\caption{The figures presented herein correspond to the phonon frequency $\omega_H/t=\omega_B/t=1.0$. The average Holstein phonon numbers (a) and the bond phonon numbers (b) as a function of bond coupling strength $g_B/t$ at fixed Holstein coupling $g_H/t=1.5$ (blue up triangles), 2.0 (green down triangles), and $2.4$ (orange diamonds). The inset is the average Holstein phonon number for the Holstein model as a function of Holstein electron-phonon coupling $g_H/t$ at $\omega_H/t=1.0$. If not visible, error bars are within the symbol size.
}
\label{FIG3}
\end{figure}

\section{Polaron with Holstein and bond electron-phonon couplings}
\label{sec:sec5}


Within this section, we study the polaron with both Holstein and bond electron-phonon couplings. We explore two distinct scenarios: (i) when the phonon frequency associated with the Holstein coupling matches that of the bond-type coupling, denoted as $\omega_H/t=\omega_B/t$, and (ii) when the phonon frequency of the Holstein coupling is twice as large as that of the bond-type coupling, indicated as $\omega_H/t=2\omega_B/t$. In both cases, within the light mass regime of the Holstein model, the introducing of bond electron-phonon coupling does not lead to a substantial increase in the effective mass compared to the bond model. However, intriguingly, we uncover a non-monotonic relationship in the effective mass as a function of bond coupling $g_B/t$ when $g_H/t$ resides within the heavy mass regime of the Holstein model.

\subsection{Polaron with both Holstein and bond couplings at $\omega_H/t=\omega_B/t$}
\label{sec:sec51}

\begin{figure}[h]
\includegraphics[width=0.41\textwidth]{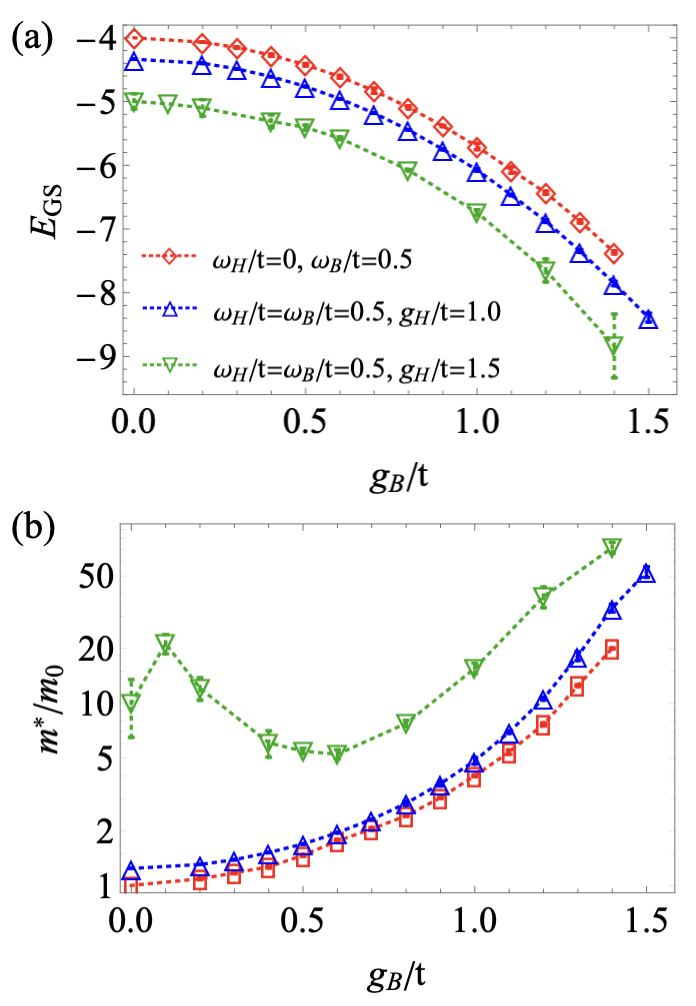}
\caption{The figures presented herein correspond to the phonon frequency $\omega_H/t=\omega_B/t=0.5$. The properties of the Holstein polaron: the ground state energy $E_{GS}$ (a), the effective mass $m^*/m_0$ (b), at electron-phonon coupling strength $g_H/t=1.0$ (blue up triangles) and 1.5 (green down triangles) as a function of the bond electron-phonon coupling $g_B/t$. The properties of the bond polaon as a function of the bond electron-phonon coupling $g_B/t$ at the bond phonon frequency $\omega_B/t=0.5$ are also presented here for comparison (red rectangles). If not visible, error bars are within symbol size. 
} 
\label{FIG4}
\end{figure}

\begin{figure}[h]
\includegraphics[width=0.39\textwidth]{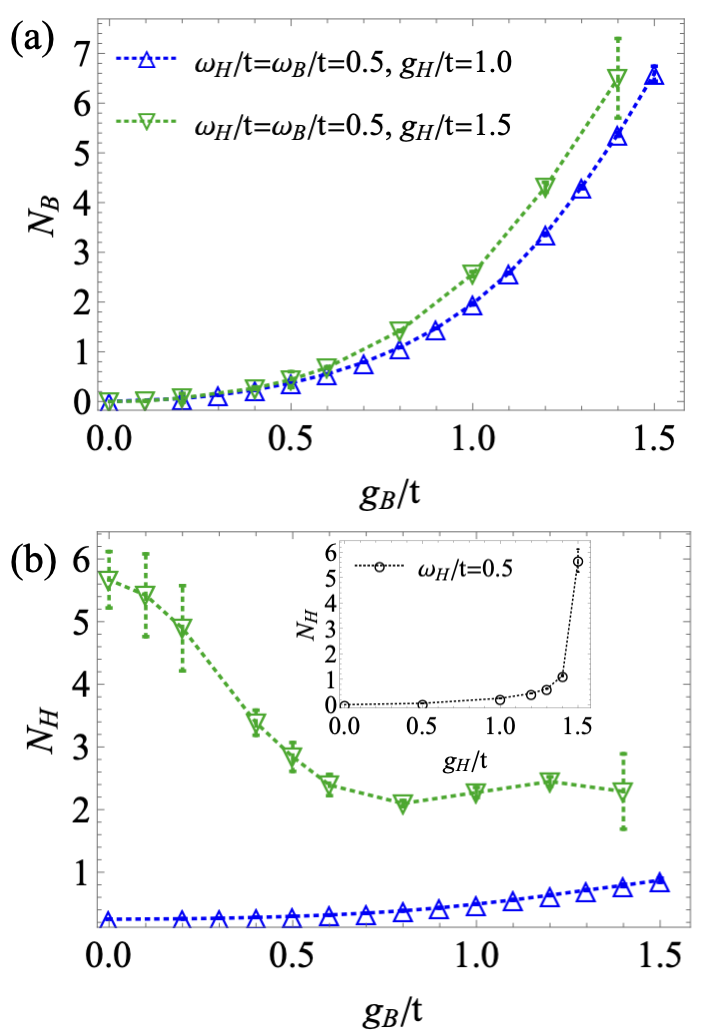}
\caption{The figures presented herein correspond to the phonon frequency $\omega_H/t=\omega_B/t=0.5$. The average Holstein phonon numbers (a) and the bond phonon numbers (b) as a function of bond coupling strength $g_B/t$ at fixed Holstein coupling $g_H/t=1.0$ (blue up triangles) and 1.5 (green down triangles). The inset is the average Holstein phonon number for the Holstein model as a function of Holstein electron-phonon coupling $g_H/t$ at $\omega_H/t=0.5$. If not visible, error bars are within the symbol size.
} 
\label{FIG5}
\end{figure}

In this section, we investigate the case where the Holstein and bond couplings feature identical phonon frequencies $\omega_H/t=\omega_B/t$. Specifically, we explore two distinct scenarios: $\omega_H/t=\omega_B/t=1.0$ and $\omega_H/t=\omega_B/t=0.5$, both are in the deep adiabatic regime. For each instance of $\omega_H/t=\omega_B/t$, the Holstein coupling parameter $g_H/t$ are fixed at several values, covering the entire range from weak to strong coupling regimes. Our analysis encompasses pivotal properties, including the ground state energy, the effective mass, as well as the average phonon numbers of Holstein and bond types, all examined as functions of the bond coupling $g_B/t$. In order to facilitate a comprehensive comparative analysis, we also present the corresponding properties of the bond model without Holstein coupling, utilizing data from Ref~\cite{PhysRevB.108.075156}.

Figure~\ref{FIG2} depicts the ground state energy $E_{GS}$ (a) and effective mass $m^*/m_0$ (b) of the Holstein polaron at fixed electron-phonon coupling strengths $g_H/t=1.5$ (blue up triangles), $2.0$ (green down triangles), and $2.4$ (orange diamonds), as well as in the absence of Holstein electron-phonon coupling $g_H/t=0$ (red rectangles). These properties are presented as functions of the bond electron-phonon coupling $g_B/t$. In comparison to the bond polaron (without Holstein coupling and $g_H/t=0$) with a phonon frequency of $\omega_B/t = 1.0$, we observe that the ground state energy $E_{GS}$ undergoes a smooth decrease as the coupling strength $g_B/t$ increases, a trend observed across all three Holstein coupling strengths ($g_H/t=1.5, 2.0,$ and $2.4$). Notably, at equivalent electron-phonon coupling strengths, the ground state energy is lower for larger Holstein coupling values.

The behavior of the effective mass $m^{*}/m_0$ exhibits a more unusual pattern, contingent upon the Holstein coupling $g_H/t$. In the light mass regime of the Holstein coupling, here $g_H/t=1.5$ and 2.0, as illustrated in Figure~\ref{FIG1}(b) with a crossover from a light to a heavy polaron state occurring at approximately $g_H/t \sim 2.2$ for the Holstein polaron with Holstein phonon frequency $\omega_H/t=1.0$, the effective mass increases with respect to the bond coupling strength $g_B/t$. This increase undergoes a change in slope as the bond coupling strength $g_B/t$ grows, observed for two Holstein coupling $g_H/t=1.5$ and $2.0$. However, intriguingly, the effective mass follows a non-monotonic trend as a function of the bond coupling $g_B/t$ when $g_H/t=2.4$ is in the heavy mass regime of the Holstein model. Initially, the effective mass decreases from around 10 to 5 as the bond coupling strength reaches approximately $g_B/t \sim 1.0$. This phenomenon indicates that the addition of bond coupling to the Holstein model within the heavy mass regime leads to a significantly lighter polaron (approximately two times lighter). Subsequently, as the bond coupling continues to increase, the effective mass begins to increase. These results underscore the existence of a light polaron in the presence of both Holstein and bond couplings at a strong coupling regime.

In Figure~\ref{FIG3}, we present the average Holstein phonon numbers (a) and average bond phonon numbers (b) as functions of the bond coupling strength $g_B/t$ for $\omega_H/t=\omega_B/t=1.0$, while keeping the Holstein coupling fixed at $g_H/t=1.5$(blue up triangles), 2.0(green down triangles), and 2.4(orange diamonds). For the average phonon number of the bond type, we observe a consistent increase with increasing bond coupling strength, regardless of the Holstein coupling. Remarkably, this quantity remains nearly constant within the range of measurement errors for varying Holstein coupling values. This suggests that the inclusion of Holstein coupling has a minimal impact on the average bond phonon number. Conversely, the average phonon number of the Holstein type exhibits significant variations across all three cases. In the light mass regime, the introducing of bond coupling has a relatively modest effect on the average Holstein phonon number. For example, with $g_H/t=1.5$, the average Holstein phonon number increases from 0 to 0.5 as a function of $g_B/t$. At $g_H/t=2.0$, which corresponds to the coupling strength around the crossover for the Holstein model, the average Holstein phonon number shows slight fluctuations, decreasing first and then increasing with respect to $g_B/t$. In the heavy mass regime with $g_H/t=2.4$, the average Holstein phonon number initially decreases from 4 to 1.2 until reaching approximately at $g_B/t \sim 1.5$. Beyond this point, it gradually starts to increase at a slower rate.


In Figure~\ref{FIG4} and Figure~\ref{FIG5}, we explore the properties of the Holstein polaron, including the ground state energy $E_{GS}$ (a) and effective mass $m^*/m_0$ (b), for electron-phonon coupling strengths $g_H/t=1.0$ and $1.5$ as a function of the bond electron-phonon coupling $g_B/t$ at $\omega_H/t=\omega_B/t=0.5$. For comparison, we also provide the ground state energy and effective mass of the bond polaron without Holstein coupling (red rectangles). The observed behavior of the ground state energy and effective mass is similar to that of the $\omega_H/t=\omega_B/t=1.0$ case. In the heavy Holstein polaron regime at $g_H/t=1.5$, the effective mass follows a more complicate non-monotonic trend, initially increasing from 10 (with no bond coupling, $g_B/t=0$) to 20 as bond coupling increases to $g_B/t=0.1$, then start decreasing from 20 to 5 at $g_B/t \sim 0.6$, rendering it approximately four times lighter. Then the effective mass begins to increase as the bond coupling strength exceeds $g_B/t \sim 0.6$. Figure~\ref{FIG5} shows that when considering a higher Holstein coupling $g_H/t=1.5$, the average number of bond phonons also has the non-monotonic behavior as a function of bond coupling $g_B/t$.

\begin{figure}[t]
\includegraphics[width=0.49\textwidth]{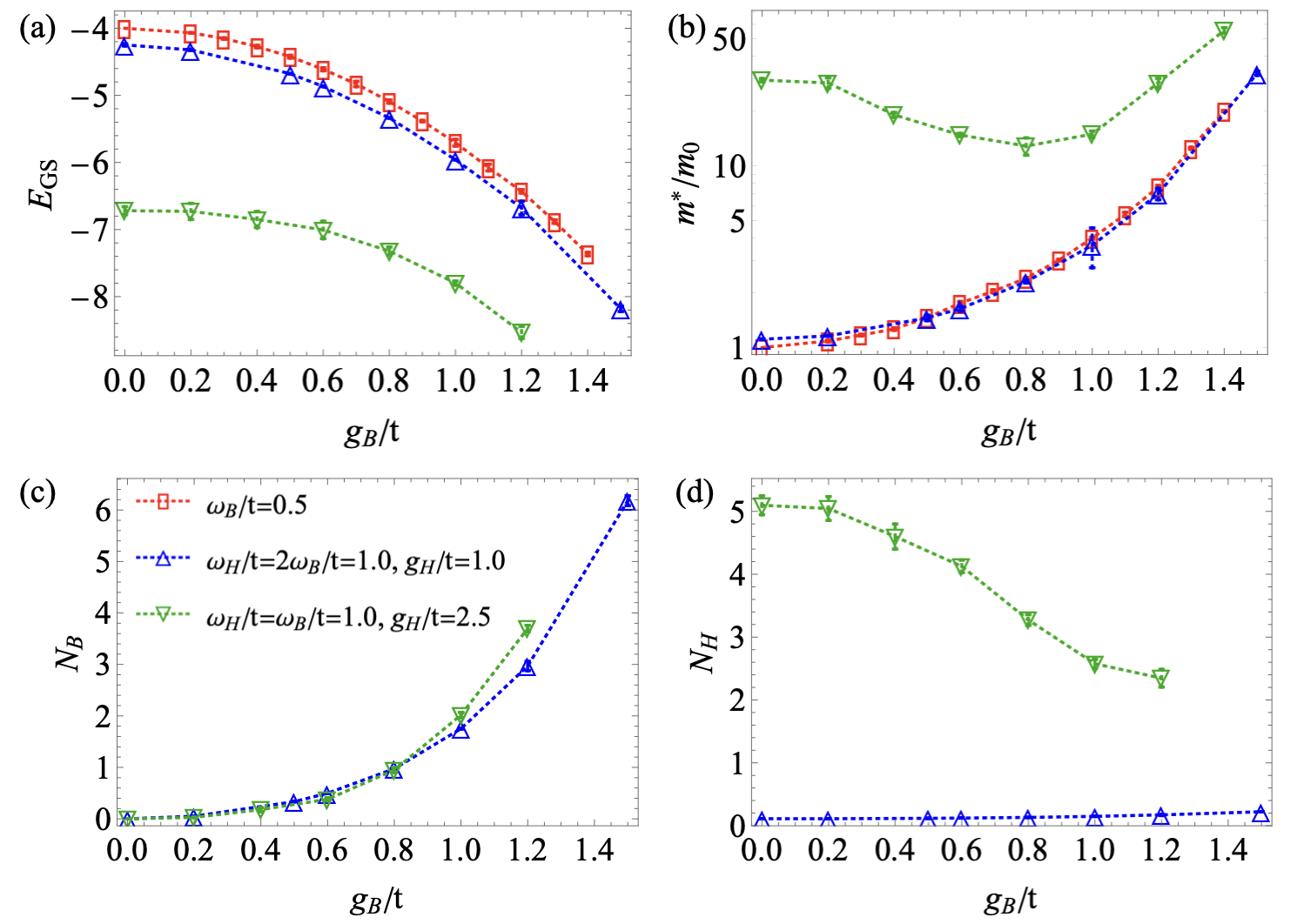}
\caption{The figures presented herein correspond to the phonon frequency $\omega_H/t=2\omega_B/t=1.0$. The properties of the Holstein polaron: ground state energy $E_{GS}$ (a), effective mass $m^*/m_0$ (b), the average phonon number of the bond type (c), and the average phonon number of Holstein type (d), at electron-phonon coupling strength $g_H/t=1.0$ (blue up triangles), and 2.5 (green down triangles) and without Holstein electron-phonon coupling $g_H/t=0$ (red rectangles) as a function of the bond electron-phonon coupling $g_B/t$. If not visible, error bars are within symbol size.
} 
\label{FIG6}
\end{figure}

\begin{figure}[t]
\includegraphics[width=0.49\textwidth]{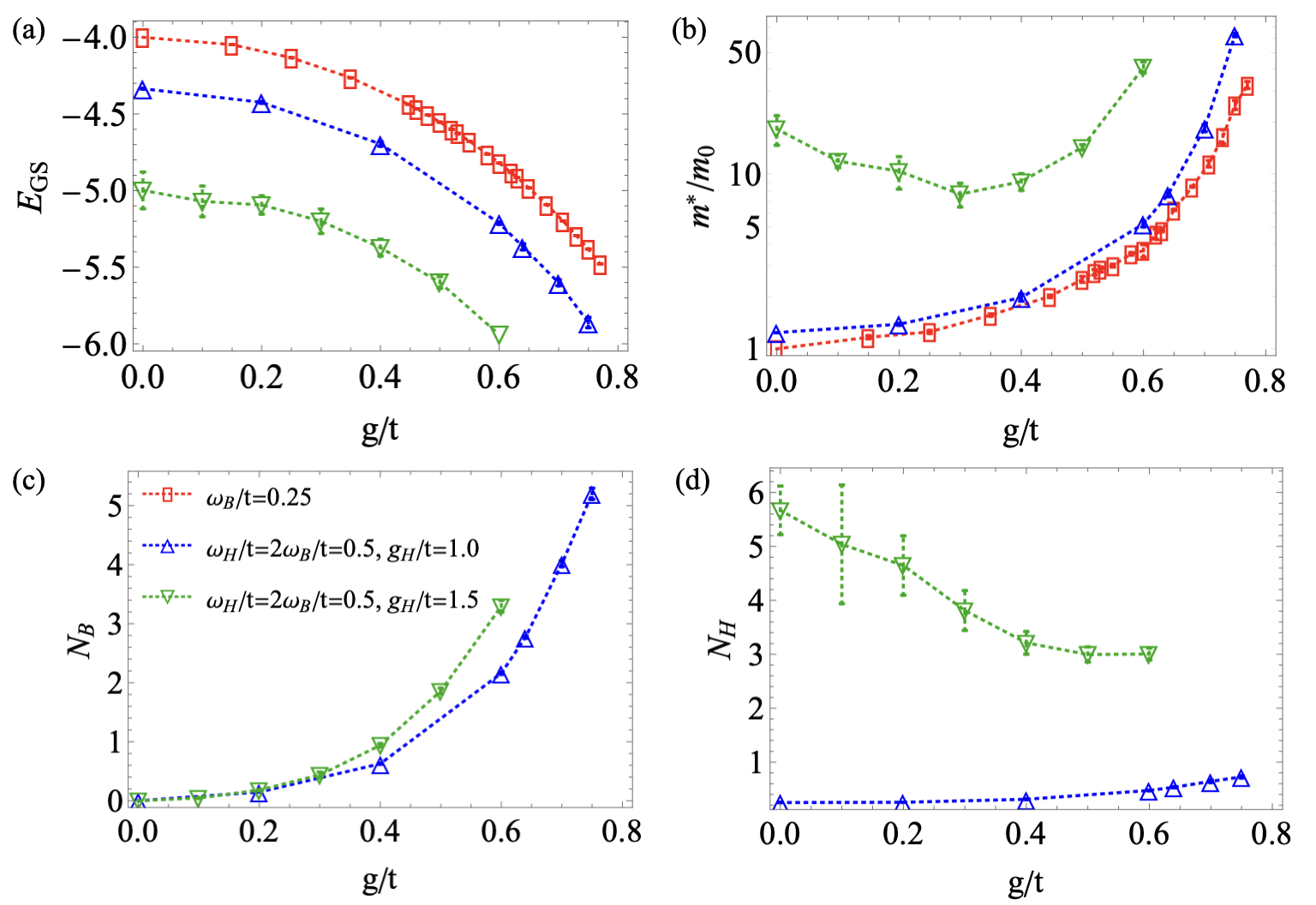}
\caption{The figures presented herein correspond to the phonon frequency $\omega_H/t=2\omega_B/t=0.5$. The properties of the Holstein polaron: ground state energy $E_{GS}$ (a), effective mass $m^*/m_0$ (b), the average phonon number of the bond type (c), and the average phonon number of Holstein type (d), at electron-phonon coupling strength $g_H/t=1.0$ (blue up triangles), 1.5 (green down triangles), and without Holstein electron-phonon coupling $g_H/t=0$ (red rectangles) as a function of the bond electron-phonon coupling $g_B/t$. If not visible, error bars are within symbol size.
} 
\label{FIG7}
\end{figure}

\subsection{Polaron with both Holstein and bond couplings at $\omega_H/t=2\omega_B/t$}
\label{sec:sec52}

In this section, we delve into scenarios featuring different phonon frequencies, namely, $\omega_H/t=2\omega_B/t$, which aligns more closely with real materials where Holstein phonons typically exhibit higher frequencies than bond-type phonons. We investigate two cases: (i) $\omega_H/t=2\omega_B/t=1.0$ and (ii) $\omega_H/t=2\omega_B/t=0.5$, both are in the adiabatic regime. 

As it can be seen from Fig.~\ref{FIG6} and Fig.~\ref{FIG7}, for both of these cases, we observe similar trends in the behavior of the ground state energy, the effective mass, and the average phonon numbers for both Holstein and bond types as in Sec.~\ref{sec:sec51}. The main difference lies in at which bond coupling the effective mass starts to have the non-monotonic behavior. For instance, in the case of $\omega_H/t=2\omega_B/t=0.5$, this non-monotonic behavior initiates at a considerably lower bond coupling strength, approximately $g_B/t \sim 0.3$, compared to the $\omega_H/t=\omega_B/t=0.5$ case, where it begins around $g_B/t \sim 0.6$.

\section{Conclusion}
\label{sec:sec6}

In conclusion, we investigate the properties of a polaron in the presence of both the Holstein and bond coupling on a two-dimensional square lattice. The properties we study here include the effective mass, ground state energy, and the number of phonons of these two types. We consider two distinct scenarios: $\omega_H/t=\omega_B/t$ and $\omega_H/t=2\omega_B/t$. We find a light polaron even within the strong electron-phonon coupling regime. What's particularly surprising is the non-monotonic relationship observed in the effective mass as a function of the strong bond electron-phonon coupling parameter, $g_B/t$, when the Holstein electron-phonon coupling parameter, $g_H/t$, falls into the heavy mass regime of the Holstein model. The non-monotonic behavior of a polaron's effective mass at strong coupling strength is recently observed in the polaron with quadratic electron-phonon interaction~\cite{ZJZhang2023}. The existence of light polarons in the presence of strong electron-phonon coupling serves as a promising starting point for the investigation of bipolarons, where large binding energies are requisite for compact size. Further exploration of bipolarons in the context of both coupling mechanisms is the next step in advancing our understanding of bipolaronic high-temperature superconductivity.


\begin{acknowledgments}

C. Z. thanks Nikolay Prokof'ev and Boris Svistunov for helpful discussion. This work is supported by the National Natural Science Foundation of China (NSFC) under Grant No. 12204173 and No. 12275263, the Innovation Program for Quantum Science and Technology (under Grant No. 2021ZD0301900), and the National Key R $\&$ D Program of China (under Grant No. 2018YFA0306501). 

\end{acknowledgments}

\bibliography{dispersive}

\begin{thebibliography}{34}%
\makeatletter
\providecommand \@ifxundefined [1]{%
 \@ifx{#1\undefined}
}%
\providecommand \@ifnum [1]{%
 \ifnum #1\expandafter \@firstoftwo
 \else \expandafter \@secondoftwo
 \fi
}%
\providecommand \@ifx [1]{%
 \ifx #1\expandafter \@firstoftwo
 \else \expandafter \@secondoftwo
 \fi
}%
\providecommand \natexlab [1]{#1}%
\providecommand \enquote  [1]{``#1''}%
\providecommand \bibnamefont  [1]{#1}%
\providecommand \bibfnamefont [1]{#1}%
\providecommand \citenamefont [1]{#1}%
\providecommand \href@noop [0]{\@secondoftwo}%
\providecommand \href [0]{\begingroup \@sanitize@url \@href}%
\providecommand \@href[1]{\@@startlink{#1}\@@href}%
\providecommand \@@href[1]{\endgroup#1\@@endlink}%
\providecommand \@sanitize@url [0]{\catcode `\\12\catcode `\$12\catcode
  `\&12\catcode `\#12\catcode `\^12\catcode `\_12\catcode `\%12\relax}%
\providecommand \@@startlink[1]{}%
\providecommand \@@endlink[0]{}%
\providecommand \url  [0]{\begingroup\@sanitize@url \@url }%
\providecommand \@url [1]{\endgroup\@href {#1}{\urlprefix }}%
\providecommand \urlprefix  [0]{URL }%
\providecommand \Eprint [0]{\href }%
\providecommand \doibase [0]{http://dx.doi.org/}%
\providecommand \selectlanguage [0]{\@gobble}%
\providecommand \bibinfo  [0]{\@secondoftwo}%
\providecommand \bibfield  [0]{\@secondoftwo}%
\providecommand \translation [1]{[#1]}%
\providecommand \BibitemOpen [0]{}%
\providecommand \bibitemStop [0]{}%
\providecommand \bibitemNoStop [0]{.\EOS\space}%
\providecommand \EOS [0]{\spacefactor3000\relax}%
\providecommand \BibitemShut  [1]{\csname bibitem#1\endcsname}%
\let\auto@bib@innerbib\@empty
\bibitem [{\citenamefont {Landau}(1933)}]{Landau33}%
  \BibitemOpen
  \bibfield  {author} {\bibinfo {author} {\bibfnamefont {L.~D.}\ \bibnamefont
  {Landau}},\ }\href@noop {} {\bibfield  {journal} {\bibinfo  {journal} {Z.
  Sowjetunion}\ }\textbf {\bibinfo {volume} {3}},\ \bibinfo {pages} {664}
  (\bibinfo {year} {1933})}\BibitemShut {NoStop}%
\bibitem [{\citenamefont {Fr\"{o}hlich}\ \emph {et~al.}(1950)\citenamefont
  {Fr\"{o}hlich}, \citenamefont {Pelzer},\ and\ \citenamefont
  {Zienau}}]{Frohlich50}%
  \BibitemOpen
  \bibfield  {author} {\bibinfo {author} {\bibfnamefont {H.}~\bibnamefont
  {Fr\"{o}hlich}}, \bibinfo {author} {\bibfnamefont {H.}~\bibnamefont
  {Pelzer}}, \ and\ \bibinfo {author} {\bibfnamefont {S.}~\bibnamefont
  {Zienau}},\ }\href@noop {} {\bibfield  {journal} {\bibinfo  {journal}
  {Philos. Mag.}\ }\textbf {\bibinfo {volume} {41}},\ \bibinfo {pages} {221}
  (\bibinfo {year} {1950})}\BibitemShut {NoStop}%
\bibitem [{\citenamefont {Feynman}(1955)}]{Feynman:1955du}%
  \BibitemOpen
  \bibfield  {author} {\bibinfo {author} {\bibfnamefont {R.~P.}\ \bibnamefont
  {Feynman}},\ }\href {\doibase 10.1103/PhysRev.97.660} {\bibfield  {journal}
  {\bibinfo  {journal} {Phys. Rev.}\ }\textbf {\bibinfo {volume} {97}},\
  \bibinfo {pages} {660} (\bibinfo {year} {1955})}\BibitemShut {NoStop}%
\bibitem [{\citenamefont {Schultz}(1959)}]{Schultz:1959el}%
  \BibitemOpen
  \bibfield  {author} {\bibinfo {author} {\bibfnamefont {T.~D.}\ \bibnamefont
  {Schultz}},\ }\href {\doibase 10.1103/PhysRev.116.526} {\bibfield  {journal}
  {\bibinfo  {journal} {Phys. Rev.}\ }\textbf {\bibinfo {volume} {116}},\
  \bibinfo {pages} {526} (\bibinfo {year} {1959})}\BibitemShut {NoStop}%
\bibitem [{\citenamefont {Holstein}(1959)}]{Holstein59}%
  \BibitemOpen
  \bibfield  {author} {\bibinfo {author} {\bibfnamefont {T.}~\bibnamefont
  {Holstein}},\ }\href@noop {} {\bibfield  {journal} {\bibinfo  {journal} {Ann.
  Phys.}\ }\textbf {\bibinfo {volume} {8}},\ \bibinfo {pages} {325} (\bibinfo
  {year} {1959})}\BibitemShut {NoStop}%
\bibitem [{\citenamefont {Alexandrov}\ and\ \citenamefont
  {Kornilovitch}(1999)}]{Alexandrov:1999fy}%
  \BibitemOpen
  \bibfield  {author} {\bibinfo {author} {\bibfnamefont {A.~S.}\ \bibnamefont
  {Alexandrov}}\ and\ \bibinfo {author} {\bibfnamefont {P.~E.}\ \bibnamefont
  {Kornilovitch}},\ }\href {\doibase 10.1103/PhysRevLett.82.807} {\bibfield
  {journal} {\bibinfo  {journal} {Phys. Rev. Lett.}\ }\textbf {\bibinfo
  {volume} {82}},\ \bibinfo {pages} {807} (\bibinfo {year} {1999})}\BibitemShut
  {NoStop}%
\bibitem [{\citenamefont {Holstein}(2000)}]{Holstein2000}%
  \BibitemOpen
  \bibfield  {author} {\bibinfo {author} {\bibfnamefont {T.}~\bibnamefont
  {Holstein}},\ }\href@noop {} {\bibfield  {journal} {\bibinfo  {journal} {Ann.
  Phys.}\ }\textbf {\bibinfo {volume} {281}},\ \bibinfo {pages} {725} (\bibinfo
  {year} {2000})}\BibitemShut {NoStop}%
\bibitem [{\citenamefont {Zhang}\ \emph
  {et~al.}(2023{\natexlab{a}})\citenamefont {Zhang}, \citenamefont {Sous},
  \citenamefont {Reichman}, \citenamefont {Berciu}, \citenamefont {Millis},
  \citenamefont {Prokof'ev},\ and\ \citenamefont
  {Svistunov}}]{PhysRevX.13.011010}%
  \BibitemOpen
  \bibfield  {author} {\bibinfo {author} {\bibfnamefont {C.}~\bibnamefont
  {Zhang}}, \bibinfo {author} {\bibfnamefont {J.}~\bibnamefont {Sous}},
  \bibinfo {author} {\bibfnamefont {D.~R.}\ \bibnamefont {Reichman}}, \bibinfo
  {author} {\bibfnamefont {M.}~\bibnamefont {Berciu}}, \bibinfo {author}
  {\bibfnamefont {A.~J.}\ \bibnamefont {Millis}}, \bibinfo {author}
  {\bibfnamefont {N.~V.}\ \bibnamefont {Prokof'ev}}, \ and\ \bibinfo {author}
  {\bibfnamefont {B.~V.}\ \bibnamefont {Svistunov}},\ }\href {\doibase
  10.1103/PhysRevX.13.011010} {\bibfield  {journal} {\bibinfo  {journal} {Phys.
  Rev. X}\ }\textbf {\bibinfo {volume} {13}},\ \bibinfo {pages} {011010}
  (\bibinfo {year} {2023}{\natexlab{a}})}\BibitemShut {NoStop}%
\bibitem [{\citenamefont {Sous}\ \emph {et~al.}(2022)\citenamefont {Sous},
  \citenamefont {Zhang}, \citenamefont {Berciu}, \citenamefont {Reichman},
  \citenamefont {Svistunov}, \citenamefont {Prokof’ev},\ and\ \citenamefont
  {Millis}}]{Johnarxiv2022}%
  \BibitemOpen
  \bibfield  {author} {\bibinfo {author} {\bibfnamefont {J.}~\bibnamefont
  {Sous}}, \bibinfo {author} {\bibfnamefont {C.}~\bibnamefont {Zhang}},
  \bibinfo {author} {\bibfnamefont {M.}~\bibnamefont {Berciu}}, \bibinfo
  {author} {\bibfnamefont {D.}~\bibnamefont {Reichman}}, \bibinfo {author}
  {\bibfnamefont {B.}~\bibnamefont {Svistunov}}, \bibinfo {author}
  {\bibfnamefont {N.}~\bibnamefont {Prokof’ev}}, \ and\ \bibinfo {author}
  {\bibfnamefont {A.}~\bibnamefont {Millis}},\ }\href {\doibase
  arXiv:2210.14236} {\bibfield  {journal} {\bibinfo  {journal}
  {arXiv:2210.14236}\ } (\bibinfo {year} {2022}),\
  arXiv:2210.14236}\BibitemShut {NoStop}%
\bibitem [{\citenamefont {Zhang}\ \emph
  {et~al.}(2023{\natexlab{b}})\citenamefont {Zhang}, \citenamefont
  {Capogrosso-Sansone}, \citenamefont {Boninsegni}, \citenamefont {Prokof'ev},\
  and\ \citenamefont {Svistunov}}]{PhysRevLett.130.236001}%
  \BibitemOpen
  \bibfield  {author} {\bibinfo {author} {\bibfnamefont {C.}~\bibnamefont
  {Zhang}}, \bibinfo {author} {\bibfnamefont {B.}~\bibnamefont
  {Capogrosso-Sansone}}, \bibinfo {author} {\bibfnamefont {M.}~\bibnamefont
  {Boninsegni}}, \bibinfo {author} {\bibfnamefont {N.~V.}\ \bibnamefont
  {Prokof'ev}}, \ and\ \bibinfo {author} {\bibfnamefont {B.~V.}\ \bibnamefont
  {Svistunov}},\ }\href {\doibase 10.1103/PhysRevLett.130.236001} {\bibfield
  {journal} {\bibinfo  {journal} {Phys. Rev. Lett.}\ }\textbf {\bibinfo
  {volume} {130}},\ \bibinfo {pages} {236001} (\bibinfo {year}
  {2023}{\natexlab{b}})}\BibitemShut {NoStop}%
\bibitem [{\citenamefont {Kornilovitch}\ and\ \citenamefont
  {Pike}(1997)}]{PhysRevB.55.R8634}%
  \BibitemOpen
  \bibfield  {author} {\bibinfo {author} {\bibfnamefont {P.~E.}\ \bibnamefont
  {Kornilovitch}}\ and\ \bibinfo {author} {\bibfnamefont {E.~R.}\ \bibnamefont
  {Pike}},\ }\href {\doibase 10.1103/PhysRevB.55.R8634} {\bibfield  {journal}
  {\bibinfo  {journal} {Phys. Rev. B}\ }\textbf {\bibinfo {volume} {55}},\
  \bibinfo {pages} {R8634} (\bibinfo {year} {1997})}\BibitemShut {NoStop}%
\bibitem [{\citenamefont {Marchand}\ \emph {et~al.}(2010)\citenamefont
  {Marchand}, \citenamefont {De~Filippis}, \citenamefont {Cataudella},
  \citenamefont {Berciu}, \citenamefont {Nagaosa}, \citenamefont {Prokof'ev},
  \citenamefont {Mishchenko},\ and\ \citenamefont
  {Stamp}}]{PhysRevLett.105.266605}%
  \BibitemOpen
  \bibfield  {author} {\bibinfo {author} {\bibfnamefont {D.~J.~J.}\
  \bibnamefont {Marchand}}, \bibinfo {author} {\bibfnamefont {G.}~\bibnamefont
  {De~Filippis}}, \bibinfo {author} {\bibfnamefont {V.}~\bibnamefont
  {Cataudella}}, \bibinfo {author} {\bibfnamefont {M.}~\bibnamefont {Berciu}},
  \bibinfo {author} {\bibfnamefont {N.}~\bibnamefont {Nagaosa}}, \bibinfo
  {author} {\bibfnamefont {N.~V.}\ \bibnamefont {Prokof'ev}}, \bibinfo {author}
  {\bibfnamefont {A.~S.}\ \bibnamefont {Mishchenko}}, \ and\ \bibinfo {author}
  {\bibfnamefont {P.~C.~E.}\ \bibnamefont {Stamp}},\ }\href {\doibase
  10.1103/PhysRevLett.105.266605} {\bibfield  {journal} {\bibinfo  {journal}
  {Phys. Rev. Lett.}\ }\textbf {\bibinfo {volume} {105}},\ \bibinfo {pages}
  {266605} (\bibinfo {year} {2010})}\BibitemShut {NoStop}%
\bibitem [{\citenamefont {Bonca}\ \emph {et~al.}(2000)\citenamefont {Bonca},
  \citenamefont {Katrasnik},\ and\ \citenamefont
  {Trugman}}]{PhysRevLett.84.3153}%
  \BibitemOpen
  \bibfield  {author} {\bibinfo {author} {\bibfnamefont {J.}~\bibnamefont
  {Bonca}}, \bibinfo {author} {\bibfnamefont {T.}~\bibnamefont {Katrasnik}}, \
  and\ \bibinfo {author} {\bibfnamefont {S.~A.}\ \bibnamefont {Trugman}},\
  }\href {\doibase 10.1103/PhysRevLett.84.3153} {\bibfield  {journal} {\bibinfo
   {journal} {Phys. Rev. Lett.}\ }\textbf {\bibinfo {volume} {84}},\ \bibinfo
  {pages} {3153} (\bibinfo {year} {2000})}\BibitemShut {NoStop}%
\bibitem [{\citenamefont {Macridin}\ \emph {et~al.}(2004)\citenamefont
  {Macridin}, \citenamefont {Sawatzky},\ and\ \citenamefont
  {Jarrell}}]{PhysRevB.69.245111}%
  \BibitemOpen
  \bibfield  {author} {\bibinfo {author} {\bibfnamefont {A.}~\bibnamefont
  {Macridin}}, \bibinfo {author} {\bibfnamefont {G.~A.}\ \bibnamefont
  {Sawatzky}}, \ and\ \bibinfo {author} {\bibfnamefont {M.}~\bibnamefont
  {Jarrell}},\ }\href {\doibase 10.1103/PhysRevB.69.245111} {\bibfield
  {journal} {\bibinfo  {journal} {Phys. Rev. B}\ }\textbf {\bibinfo {volume}
  {69}},\ \bibinfo {pages} {245111} (\bibinfo {year} {2004})}\BibitemShut
  {NoStop}%
\bibitem [{\citenamefont {Chakraverty}\ \emph {et~al.}(1998)\citenamefont
  {Chakraverty}, \citenamefont {Ranninger},\ and\ \citenamefont
  {Feinberg}}]{PhysRevLett.81.433}%
  \BibitemOpen
  \bibfield  {author} {\bibinfo {author} {\bibfnamefont {B.~K.}\ \bibnamefont
  {Chakraverty}}, \bibinfo {author} {\bibfnamefont {J.}~\bibnamefont
  {Ranninger}}, \ and\ \bibinfo {author} {\bibfnamefont {D.}~\bibnamefont
  {Feinberg}},\ }\href {\doibase 10.1103/PhysRevLett.81.433} {\bibfield
  {journal} {\bibinfo  {journal} {Phys. Rev. Lett.}\ }\textbf {\bibinfo
  {volume} {81}},\ \bibinfo {pages} {433} (\bibinfo {year} {1998})}\BibitemShut
  {NoStop}%
\bibitem [{\citenamefont {Verbist}\ \emph {et~al.}(1991)\citenamefont
  {Verbist}, \citenamefont {Peeters},\ and\ \citenamefont
  {Devreese}}]{PhysRevB.43.2712}%
  \BibitemOpen
  \bibfield  {author} {\bibinfo {author} {\bibfnamefont {G.}~\bibnamefont
  {Verbist}}, \bibinfo {author} {\bibfnamefont {F.~M.}\ \bibnamefont
  {Peeters}}, \ and\ \bibinfo {author} {\bibfnamefont {J.~T.}\ \bibnamefont
  {Devreese}},\ }\href {\doibase 10.1103/PhysRevB.43.2712} {\bibfield
  {journal} {\bibinfo  {journal} {Phys. Rev. B}\ }\textbf {\bibinfo {volume}
  {43}},\ \bibinfo {pages} {2712} (\bibinfo {year} {1991})}\BibitemShut
  {NoStop}%
\bibitem [{\citenamefont {da~Costa}\ and\ \citenamefont
  {Peeters}(1996)}]{Costa_1996}%
  \BibitemOpen
  \bibfield  {author} {\bibinfo {author} {\bibfnamefont {W.~B.}\ \bibnamefont
  {da~Costa}}\ and\ \bibinfo {author} {\bibfnamefont {F.~M.}\ \bibnamefont
  {Peeters}},\ }\href {\doibase 10.1088/0953-8984/8/13/009} {\bibfield
  {journal} {\bibinfo  {journal} {Journal of Physics: Condensed Matter}\
  }\textbf {\bibinfo {volume} {8}},\ \bibinfo {pages} {2173} (\bibinfo {year}
  {1996})}\BibitemShut {NoStop}%
\bibitem [{\citenamefont {Farias}\ \emph {et~al.}(1996)\citenamefont {Farias},
  \citenamefont {da~Costa},\ and\ \citenamefont {Peeters}}]{PhysRevB.54.12835}%
  \BibitemOpen
  \bibfield  {author} {\bibinfo {author} {\bibfnamefont {G.~A.}\ \bibnamefont
  {Farias}}, \bibinfo {author} {\bibfnamefont {W.~B.}\ \bibnamefont
  {da~Costa}}, \ and\ \bibinfo {author} {\bibfnamefont {F.~M.}\ \bibnamefont
  {Peeters}},\ }\href {\doibase 10.1103/PhysRevB.54.12835} {\bibfield
  {journal} {\bibinfo  {journal} {Phys. Rev. B}\ }\textbf {\bibinfo {volume}
  {54}},\ \bibinfo {pages} {12835} (\bibinfo {year} {1996})}\BibitemShut
  {NoStop}%
\bibitem [{\citenamefont {Zhang}\ \emph {et~al.}(2021)\citenamefont {Zhang},
  \citenamefont {Prokof'ev},\ and\ \citenamefont
  {Svistunov}}]{PhysRevB.104.035143}%
  \BibitemOpen
  \bibfield  {author} {\bibinfo {author} {\bibfnamefont {C.}~\bibnamefont
  {Zhang}}, \bibinfo {author} {\bibfnamefont {N.~V.}\ \bibnamefont
  {Prokof'ev}}, \ and\ \bibinfo {author} {\bibfnamefont {B.~V.}\ \bibnamefont
  {Svistunov}},\ }\href {\doibase 10.1103/PhysRevB.104.035143} {\bibfield
  {journal} {\bibinfo  {journal} {Phys. Rev. B}\ }\textbf {\bibinfo {volume}
  {104}},\ \bibinfo {pages} {035143} (\bibinfo {year} {2021})}\BibitemShut
  {NoStop}%
\bibitem [{\citenamefont {Carbone}\ \emph {et~al.}(2021)\citenamefont
  {Carbone}, \citenamefont {Millis}, \citenamefont {Reichman},\ and\
  \citenamefont {Sous}}]{PhysRevB.104.L140307}%
  \BibitemOpen
  \bibfield  {author} {\bibinfo {author} {\bibfnamefont {M.~R.}\ \bibnamefont
  {Carbone}}, \bibinfo {author} {\bibfnamefont {A.~J.}\ \bibnamefont {Millis}},
  \bibinfo {author} {\bibfnamefont {D.~R.}\ \bibnamefont {Reichman}}, \ and\
  \bibinfo {author} {\bibfnamefont {J.}~\bibnamefont {Sous}},\ }\href {\doibase
  10.1103/PhysRevB.104.L140307} {\bibfield  {journal} {\bibinfo  {journal}
  {Phys. Rev. B}\ }\textbf {\bibinfo {volume} {104}},\ \bibinfo {pages}
  {L140307} (\bibinfo {year} {2021})}\BibitemShut {NoStop}%
\bibitem [{\citenamefont {Zhang}\ \emph {et~al.}(2022)\citenamefont {Zhang},
  \citenamefont {Prokof'ev},\ and\ \citenamefont
  {Svistunov}}]{PhysRevB.105.L020501}%
  \BibitemOpen
  \bibfield  {author} {\bibinfo {author} {\bibfnamefont {C.}~\bibnamefont
  {Zhang}}, \bibinfo {author} {\bibfnamefont {N.~V.}\ \bibnamefont
  {Prokof'ev}}, \ and\ \bibinfo {author} {\bibfnamefont {B.~V.}\ \bibnamefont
  {Svistunov}},\ }\href {\doibase 10.1103/PhysRevB.105.L020501} {\bibfield
  {journal} {\bibinfo  {journal} {Phys. Rev. B}\ }\textbf {\bibinfo {volume}
  {105}},\ \bibinfo {pages} {L020501} (\bibinfo {year} {2022})}\BibitemShut
  {NoStop}%
\bibitem [{\citenamefont {Ku}\ \emph {et~al.}(2002)\citenamefont {Ku},
  \citenamefont {Trugman},\ and\ \citenamefont {Bon\ifmmode~\check{c}\else
  \v{c}\fi{}a}}]{PhysRevB.65.174306}%
  \BibitemOpen
  \bibfield  {author} {\bibinfo {author} {\bibfnamefont {L.-C.}\ \bibnamefont
  {Ku}}, \bibinfo {author} {\bibfnamefont {S.~A.}\ \bibnamefont {Trugman}}, \
  and\ \bibinfo {author} {\bibfnamefont {J.}~\bibnamefont
  {Bon\ifmmode~\check{c}\else \v{c}\fi{}a}},\ }\href {\doibase
  10.1103/PhysRevB.65.174306} {\bibfield  {journal} {\bibinfo  {journal} {Phys.
  Rev. B}\ }\textbf {\bibinfo {volume} {65}},\ \bibinfo {pages} {174306}
  (\bibinfo {year} {2002})}\BibitemShut {NoStop}%
\bibitem [{\citenamefont {Bon\ifmmode~\check{c}\else \v{c}\fi{}a}\ \emph
  {et~al.}(1999)\citenamefont {Bon\ifmmode~\check{c}\else \v{c}\fi{}a},
  \citenamefont {Trugman},\ and\ \citenamefont {Batisti\ifmmode~\acute{c}\else
  \'{c}\fi{}}}]{PhysRevB.60.1633}%
  \BibitemOpen
  \bibfield  {author} {\bibinfo {author} {\bibfnamefont {J.}~\bibnamefont
  {Bon\ifmmode~\check{c}\else \v{c}\fi{}a}}, \bibinfo {author} {\bibfnamefont
  {S.~A.}\ \bibnamefont {Trugman}}, \ and\ \bibinfo {author} {\bibfnamefont
  {I.}~\bibnamefont {Batisti\ifmmode~\acute{c}\else \'{c}\fi{}}},\ }\href
  {\doibase 10.1103/PhysRevB.60.1633} {\bibfield  {journal} {\bibinfo
  {journal} {Phys. Rev. B}\ }\textbf {\bibinfo {volume} {60}},\ \bibinfo
  {pages} {1633} (\bibinfo {year} {1999})}\BibitemShut {NoStop}%
\bibitem [{\citenamefont {Su}\ \emph {et~al.}(1979)\citenamefont {Su},
  \citenamefont {Schrieffer},\ and\ \citenamefont
  {Heeger}}]{PhysRevLett.42.1698}%
  \BibitemOpen
  \bibfield  {author} {\bibinfo {author} {\bibfnamefont {W.~P.}\ \bibnamefont
  {Su}}, \bibinfo {author} {\bibfnamefont {J.~R.}\ \bibnamefont {Schrieffer}},
  \ and\ \bibinfo {author} {\bibfnamefont {A.~J.}\ \bibnamefont {Heeger}},\
  }\href {\doibase 10.1103/PhysRevLett.42.1698} {\bibfield  {journal} {\bibinfo
   {journal} {Phys. Rev. Lett.}\ }\textbf {\bibinfo {volume} {42}},\ \bibinfo
  {pages} {1698} (\bibinfo {year} {1979})}\BibitemShut {NoStop}%
\bibitem [{\citenamefont {Bari\ifmmode \check{s}\else
  \v{s}\fi{}i\ifmmode~\acute{c}\else \'{c}\fi{}}\ \emph
  {et~al.}(1970)\citenamefont {Bari\ifmmode \check{s}\else
  \v{s}\fi{}i\ifmmode~\acute{c}\else \'{c}\fi{}}, \citenamefont {Labb\'e},\
  and\ \citenamefont {Friedel}}]{PhysRevLett.25.919}%
  \BibitemOpen
  \bibfield  {author} {\bibinfo {author} {\bibfnamefont {S.}~\bibnamefont
  {Bari\ifmmode \check{s}\else \v{s}\fi{}i\ifmmode~\acute{c}\else \'{c}\fi{}}},
  \bibinfo {author} {\bibfnamefont {J.}~\bibnamefont {Labb\'e}}, \ and\
  \bibinfo {author} {\bibfnamefont {J.}~\bibnamefont {Friedel}},\ }\href
  {\doibase 10.1103/PhysRevLett.25.919} {\bibfield  {journal} {\bibinfo
  {journal} {Phys. Rev. Lett.}\ }\textbf {\bibinfo {volume} {25}},\ \bibinfo
  {pages} {919} (\bibinfo {year} {1970})}\BibitemShut {NoStop}%
\bibitem [{\citenamefont {Bari\ifmmode \check{s}\else
  \v{s}\fi{}i\ifmmode~\acute{c}\else
  \'{c}\fi{}}(1972{\natexlab{a}})}]{PhysRevB.5.932}%
  \BibitemOpen
  \bibfield  {author} {\bibinfo {author} {\bibfnamefont {S.}~\bibnamefont
  {Bari\ifmmode \check{s}\else \v{s}\fi{}i\ifmmode~\acute{c}\else
  \'{c}\fi{}}},\ }\href {\doibase 10.1103/PhysRevB.5.932} {\bibfield  {journal}
  {\bibinfo  {journal} {Phys. Rev. B}\ }\textbf {\bibinfo {volume} {5}},\
  \bibinfo {pages} {932} (\bibinfo {year} {1972}{\natexlab{a}})}\BibitemShut
  {NoStop}%
\bibitem [{\citenamefont {Bari\ifmmode \check{s}\else
  \v{s}\fi{}i\ifmmode~\acute{c}\else
  \'{c}\fi{}}(1972{\natexlab{b}})}]{PhysRevB.5.941}%
  \BibitemOpen
  \bibfield  {author} {\bibinfo {author} {\bibfnamefont {S.}~\bibnamefont
  {Bari\ifmmode \check{s}\else \v{s}\fi{}i\ifmmode~\acute{c}\else
  \'{c}\fi{}}},\ }\href {\doibase 10.1103/PhysRevB.5.941} {\bibfield  {journal}
  {\bibinfo  {journal} {Phys. Rev. B}\ }\textbf {\bibinfo {volume} {5}},\
  \bibinfo {pages} {941} (\bibinfo {year} {1972}{\natexlab{b}})}\BibitemShut
  {NoStop}%
\bibitem [{\citenamefont {Prokof'ev}\ and\ \citenamefont
  {Svistunov}(1998)}]{PhysRevLett.81.2514}%
  \BibitemOpen
  \bibfield  {author} {\bibinfo {author} {\bibfnamefont {N.~V.}\ \bibnamefont
  {Prokof'ev}}\ and\ \bibinfo {author} {\bibfnamefont {B.~V.}\ \bibnamefont
  {Svistunov}},\ }\href {\doibase 10.1103/PhysRevLett.81.2514} {\bibfield
  {journal} {\bibinfo  {journal} {Phys. Rev. Lett.}\ }\textbf {\bibinfo
  {volume} {81}},\ \bibinfo {pages} {2514} (\bibinfo {year}
  {1998})}\BibitemShut {NoStop}%
\bibitem [{\citenamefont {Mishchenko}\ \emph {et~al.}(2000)\citenamefont
  {Mishchenko}, \citenamefont {Prokof'ev}, \citenamefont {Sakamoto},\ and\
  \citenamefont {Svistunov}}]{PhysRevB.62.6317}%
  \BibitemOpen
  \bibfield  {author} {\bibinfo {author} {\bibfnamefont {A.~S.}\ \bibnamefont
  {Mishchenko}}, \bibinfo {author} {\bibfnamefont {N.~V.}\ \bibnamefont
  {Prokof'ev}}, \bibinfo {author} {\bibfnamefont {A.}~\bibnamefont {Sakamoto}},
  \ and\ \bibinfo {author} {\bibfnamefont {B.~V.}\ \bibnamefont {Svistunov}},\
  }\href {\doibase 10.1103/PhysRevB.62.6317} {\bibfield  {journal} {\bibinfo
  {journal} {Phys. Rev. B}\ }\textbf {\bibinfo {volume} {62}},\ \bibinfo
  {pages} {6317} (\bibinfo {year} {2000})}\BibitemShut {NoStop}%
\bibitem [{\citenamefont {Mishchenko}\ \emph {et~al.}(2019)\citenamefont
  {Mishchenko}, \citenamefont {Pollet}, \citenamefont {Prokof'ev},
  \citenamefont {Kumar}, \citenamefont {Maslov},\ and\ \citenamefont
  {Nagaosa}}]{PhysRevLett.123.076601}%
  \BibitemOpen
  \bibfield  {author} {\bibinfo {author} {\bibfnamefont {A.~S.}\ \bibnamefont
  {Mishchenko}}, \bibinfo {author} {\bibfnamefont {L.}~\bibnamefont {Pollet}},
  \bibinfo {author} {\bibfnamefont {N.~V.}\ \bibnamefont {Prokof'ev}}, \bibinfo
  {author} {\bibfnamefont {A.}~\bibnamefont {Kumar}}, \bibinfo {author}
  {\bibfnamefont {D.~L.}\ \bibnamefont {Maslov}}, \ and\ \bibinfo {author}
  {\bibfnamefont {N.}~\bibnamefont {Nagaosa}},\ }\href {\doibase
  10.1103/PhysRevLett.123.076601} {\bibfield  {journal} {\bibinfo  {journal}
  {Phys. Rev. Lett.}\ }\textbf {\bibinfo {volume} {123}},\ \bibinfo {pages}
  {076601} (\bibinfo {year} {2019})}\BibitemShut {NoStop}%
\bibitem [{\citenamefont {Mishchenko}\ \emph {et~al.}(2014)\citenamefont
  {Mishchenko}, \citenamefont {Nagaosa},\ and\ \citenamefont
  {Prokof'ev}}]{PhysRevLett.113.166402}%
  \BibitemOpen
  \bibfield  {author} {\bibinfo {author} {\bibfnamefont {A.~S.}\ \bibnamefont
  {Mishchenko}}, \bibinfo {author} {\bibfnamefont {N.}~\bibnamefont {Nagaosa}},
  \ and\ \bibinfo {author} {\bibfnamefont {N.}~\bibnamefont {Prokof'ev}},\
  }\href {\doibase 10.1103/PhysRevLett.113.166402} {\bibfield  {journal}
  {\bibinfo  {journal} {Phys. Rev. Lett.}\ }\textbf {\bibinfo {volume} {113}},\
  \bibinfo {pages} {166402} (\bibinfo {year} {2014})}\BibitemShut {NoStop}%
\bibitem [{\citenamefont {Prokof'ev}\ and\ \citenamefont
  {Svistunov}(2022)}]{PhysRevB.106.L041117}%
  \BibitemOpen
  \bibfield  {author} {\bibinfo {author} {\bibfnamefont {N.~V.}\ \bibnamefont
  {Prokof'ev}}\ and\ \bibinfo {author} {\bibfnamefont {B.~V.}\ \bibnamefont
  {Svistunov}},\ }\href {\doibase 10.1103/PhysRevB.106.L041117} {\bibfield
  {journal} {\bibinfo  {journal} {Phys. Rev. B}\ }\textbf {\bibinfo {volume}
  {106}},\ \bibinfo {pages} {L041117} (\bibinfo {year} {2022})}\BibitemShut
  {NoStop}%
\bibitem [{\citenamefont {Zhang}(2023)}]{PhysRevB.108.075156}%
  \BibitemOpen
  \bibfield  {author} {\bibinfo {author} {\bibfnamefont {C.}~\bibnamefont
  {Zhang}},\ }\href {\doibase 10.1103/PhysRevB.108.075156} {\bibfield
  {journal} {\bibinfo  {journal} {Phys. Rev. B}\ }\textbf {\bibinfo {volume}
  {108}},\ \bibinfo {pages} {075156} (\bibinfo {year} {2023})}\BibitemShut
  {NoStop}%
\bibitem [{\citenamefont {Zhang}\ \emph
  {et~al.}(2023{\natexlab{c}})\citenamefont {Zhang}, \citenamefont {Kuklov},
  \citenamefont {Prokof’ev},\ and\ \citenamefont {Svistunov}}]{ZJZhang2023}%
  \BibitemOpen
  \bibfield  {author} {\bibinfo {author} {\bibfnamefont {Z.}~\bibnamefont
  {Zhang}}, \bibinfo {author} {\bibfnamefont {A.}~\bibnamefont {Kuklov}},
  \bibinfo {author} {\bibfnamefont {N.}~\bibnamefont {Prokof’ev}}, \ and\
  \bibinfo {author} {\bibfnamefont {B.}~\bibnamefont {Svistunov}},\ }\href
  {\doibase arXiv:2309.10669} {\bibfield  {journal} {\bibinfo  {journal}
  {arXiv:2309.10669}\ } (\bibinfo {year} {2023}{\natexlab{c}}),\
  arXiv:2309.10669}\BibitemShut {NoStop}%
\end{thebibliography}%

\end{document}